# Charge separation dynamics and opto-electronic properties of a diaminoterephthalate- $C_{60}$ diad

*Stefano Pittalis, Alain Delgado, Jörg Robin, Lena Freimuth, Jens Christoffers, Christoph Lienau, Carlo Andrea Rozzi\**

Dr. Stefano Pittalis, Dr. Carlo Andrea Rozzi, Istituto Nanoscienze – CNR, via Campi 213a, 41125 Modena, Italy
E-mail: carloandrea.rozzi@nano.cnr.it
Dr. Alain Delgado, Istituto Nanoscienze – CNR, via Campi 213a, 41125 Modena, Italy and Centro de Aplicaciones Tecnológicas y Desarrollo Nuclear, Calle 30 # 502, 11300, La Habana, Cuba.
Jörg Robin, Prof. Dr. Christoph Lienau, Institut für Physik and Center of Interface Science, Carl von Ossietzky Universität, 26111 Oldenburg, Germany
Lena Freimuth, Prof. Dr. Jens Christoffers, Institut für Chemie and Center of Interface Science, Carl von Ossietzky Universität, 26111 Oldenburg, Germany

A novel diad composed of a diaminoterephthalate scaffold, covalently linked to a Fullerene derivative, is explored as a nanosized charge separation unit powered by solar energy. Its opto-electronic properties are studied and the charge separation rate is determined. Simulations of the coupled electronic and nuclear dynamics in the Ehrenfest approximation are carried out ona sub 100 fs time scale after photoexcitation in order to gain insights about the mechanisms driving the the charge separation. In particular, the role of vibronic coupling and of the detailed morphology are highlighted.

## 1. Introduction

The quest for new materials for clean energy production and storage is one of the most compelling research fields of our times. The challenge of understanding fundamental processes in natural and man-made systems has an immediate potential for societal benefits, and is therefore of primary importance both from the scientific and from the technological point of view.

During the last twenty years there has been a clear focus shift of frontier research from the study of solid-state crystalline semiconductor devices to the investigation of new classes of devices based on organic or hybrid components. This breakthrough was first driven by the paradigm change introduced by the invention of dye-sensitized[1] and bulk heterojunction organic solar cells[2].

Solar cells based on organic semiconductors can be cheaply manufactured with low-temperature, high-throughput techniques that are not available for inorganic semiconductors. Besides, they offer much larger flexibility in the chemical manipulation and tailoring of the electronic properties. In these types of cells the primary neutral excitation (or exciton) is induced by sun light in an organic semiconductor of suitable bandgap (usually a dye or a π-conjugated polymer). The excitation energy must then be efficiently used to generate a flow of free charge carriers, for the device to work as a photovoltaic unit. At variance with first generation cells based on doped silicon, or III-V semiconductor junctions, in which the driving force for the charge separation is the intrinsic electric field generated by charge depletion across the junction at the equilibrium, organic devices have significantly smaller dielectric constant, and therefore much higher exciton binding energies. For this reason their operation relies on different mechanisms for the charge separation to be performed effectively. To date, big efforts have been spent in designing and synthesising molecular systems made of electron donor and acceptor moieties, both to mimic photosynthetic reaction centres, and to improve the performances of existing photovoltaic architectures by chemically engineering their structures. In this quest, the often adopted trial and error optimization strategies must be supported by a detailed understanding of how the photoinduced charge separation actually proceeds at a microscopic level. Most of the natural systems which are a source of inspiration are too complex to provide practical guidelines for artificial devices.

In order to understand the fundamental processes, it is crucial to consider the simplest and smallest possible system that show the same behavior as natural reaction centers (or the equivalent for photovoltaic effect).[3,4] This line of thought was followed for example by groups whose pioneering works lead to the creation of supramolecular donor-acceptor or donor-bridge-acceptor compounds.[5,6,7] These systems were so far the most elementary units for which simple analytical models could be checked against fully *ab initio* theories and simulations. Moreover recent 2D optical spectroscopy observations reinforced the idea that quantum physics could seriously affect the efficiency of energy transfer in those systems.[8] In recent works some us have studied the quantum aspects of light-driven charge dynamics in such systems on a sub 100 femtosecond (fs) time scale after photoexcitation.[9,10] Both in the case of a covalently linked caroteno-porphyrin-$C_{60}$ triad, and in the case of the P3HT:PCBM blend we found experimental evidences of quantum coherent behavior, and we could provide first principle theoretical support that the strong coupling of the electrons and nuclei was responsible for driving wavelike oscillations in the electronic system.

However both the synthesis and the accurate simulation of such arrays requires significant effort. The need for more straightforward synthesis protocols and for more elementary model systems has prompted us to seek for smaller and more functionally flexible compounds. Having at hand a modular unit would allow us to build from scratch a series of nanosized "all in one" prototypical photovoltaic cells and to control their structural details according to the desired effects on the overall functionality of the device.

For this purpose we turn our attention to a recently developed organic scaffold molecule (diaminoterephthalate) originally meant for combinatorial chemistry.[11]

Diaminoterephthalates have four sites for binding up to four different effector groups and, therefore, are suitable for different applications in biochemistry, physics and physical chemistry. They can, for example, be designed as "turn-on" fluorescent probes and were employed as tools for investigating conformational changes of neuronal proteins in dependence of $Ca^{2+}$-concentrations.[13] The scaffold can also be conveniently bound to electron donors, acceptors, gold or glass surfaces, or proteins.

In this paper we propose to attach a diaminoterephthalate unit to $C_{60}$ fullerene by a pyrrolidine linker in order to form a donor-acceptor diad. By means of theoretical computations, we address the question if this structure can be a useful building block for a molecular photovoltaic nano-sized unit. Our methods of choice are Density Functional Theory (DFT) and its time-dependent generalization (TDDFT) together with the Ehrenfest path method for nuclear dynamics.[14,15] Our investigation involves three main steps: (i) determining the stable structures of the system and its components (see Sec 2.1); (ii) determining their optical absorption spectra and identifying the chromophore excitations (see Sec 2.2); (iii) exciting the ground state of the diad and propagating its quantum state in time to seek evidences of charge transfer from the donor (chromophore) to the acceptor (fullerene) (see Sec. 2.3).

## 2. Results and discussion

### 2.1 Building blocks

In the present work we study the compound depicted in **Figure 1a**, simply referenced as "diad" along the paper. This diad is obtained by connecting a diaminoterephthalate molecule (shortly addressed as "chromophore", shown in Figure 1c) to an electron acceptor, such as one of the functionalized variants on the Fullerene theme. In our case we have chosen the pyrrolidine-$C_{60}$ unit depicted in Figure 1d. As anticipated in the previous section the chosen chromophore is particularly convenient as it can be easily functionalized with up to four different groups. For instance, thiol or silane groups can be used for binding it to gold or glass surfaces, and biomolecules could be immobilized at the surface with an alkyne or azide moiety, allowing complete control on the local environment. The pristine chromophore in Figure 1c is a colored, fluorescent compound. Its optical properties can be controlled by appropriate substituents. For example, a reaction with thiols can be designed such that a non-fluorescent species is turned into a fluorescent one which emits orange light with a quantum yield of 80%.[13] The length and conjugation of the bridge connecting the donor and the acceptor may be tailored as well.[16] Those factors are well known to affect the transfer rate, as demonstrated, e.g., by measurements performed on donor-bridge-acceptor systems built from tetracene (donor), pyromellitimide (acceptor), and oligophenylenevinylenes (bridge). These measurements show that upon selective photoexcitation of the donor, the electron-transfer rate first drops abruptly when the size of the bridge is increased, and then starts growing again to reach much higher values for longer bridge chains.[17]

The synthetic strategy for the title compound is briefly resumed in Figure 1b: Catalytic hydrogenation would yield the free mono-carboxylic acid which could then be coupled with benzylic alcohol bearing a dimethylacetal-protected aldehyde function in its para-position. After deprotection, the liberated benzaldehyde function is submitted to 1,3-dipolar cycloaddition with fullerene $C_{60}$ and sarcosine (*i.e. N*-methylglycine)[12] forming an intermediate azomethine-ylide and subsequently the pyrrolidine ring (after decarboxylation) as the linker between benzyl ester and $C_{60}$. However, the synthesis depicted in Figure 1b was not yet accomplished. This path could be adapted to allow us to optimize the linker. The length and conjugation of the "bridge" unit could then be engineered to vary the length of the electron path, to control the amount of electronic coupling between the donor and the acceptor, and to investigate the role of different channels for charge separation – such as through bond hopping as opposed to through space tunneling.

Here, we consider the shortest bridging conjugate group, i.e., a *para*-phenylene unit. The optimized geometries of the diad and of its isolated moieties have been calculated in gas phase with DFT by using the hybrid B3LYP exchange-correlation functional[19] and the standard Pople basis set 6-31G(d,p).[20] The selected diad geometry corresponds to an absolute minimum of the energy in the calculations, but the nuclear configuration shown in Figure 1a is very close to a set of degenerate arrangements involving free rotations around the C-C bond connecting the chromophore to the benzyl group, and around the contiguous C-O single bond.

Indeed, the total ground state energy difference with respect to a configuration in which the chromophore moiety is rotated 180° around the C-C bond (see **Figure 5**) is only −0.014 eV. We therefore expect all those configurations to be simultaneously active in a room-temperature solution sample. The relaxed geometries of the diad components, i.e., a pyrrolidine-$C_{60}$ acceptor, and the diaminoterephthalate moiety are also shown in Figure 1d and 1c, respectively. In the following section, we determine and compare the electronic properties of the diad with those of its isolated moieties.

## 2.2 Electronic structure and optical properties

In order to address the possible excitation pattern of the diad we have first calculated the ground state optical response of the chromophore alone. In **Figure 2** we show the absorption spectra of the chromophore (see Figure 1 (c) for the structure) calculated within the linear response time-dependent DFT (LR-TDDFT)[21] at the B3LYP level in vacuo and in a dichloromethane ($CH_2Cl_2$) solution. Solvation effects in the ground and excited states are accounted for in the framework of the conductor-like Polarizable Continuum Model (C-PCM).[22, 23] The PCM formalism allows us to map the intractable problem of considering explicitly the $CH_2Cl_2$ solvent molecules into a problem in which the chromophore interacts electrostatically with an infinite polarizable medium characterized by its static ($\varepsilon_s = 8.93$) and high-frequency ($\varepsilon_\infty = 2.02$) dielectric constants.[24] The absorption spectra are obtained by computing the excitation energies (poles of the density-density response function) and the oscillator strengths of the first twenty excited states of the chromophore. The main absorption features of the spectrum of the chromophore are also very well described at the LDA level. Of course minor quantitative differences can be spotted by closely comparing the LDA spectra (see red line in **Figure 4**) with the ones obtained with the B3LYP functional. In particular, the LDA spectra are red shifted, as expected, indicating the well-known tendency of LDA to underestimate long-range exchange-correlation energies. However, both theory levels agree that the absorption edge is due to a predominant pure transition between the highest occupied (HOMO) and the lowest unoccupied (LUMO) molecular orbitals.

Both in vacuo and in solution the calculations show two distinct absorption bands, one in the visible region and the other in the ultraviolet range. The main features of the absorption, for the solvated molecule, are centered at 464 nm (2.67 eV), 239 nm (5.18 eV) and 210 nm (6.1 eV) and are slightly red shifted by about 0.1 eV compared to the gas phase results, as it was expected given the a non-polarity of $CH_2Cl_2$. On the other hand, we notice that the solvent polarization increases the intensity of all the absorption resonances. The optical absorption onset at 464 nm corresponds to the lowest-lying excited state showing a dominant excitation amplitude for the transition between the HOMO and LUMO. The isosurfaces of these orbitals responsible for the absorption in the visible appear in the inset of Figure 2. The chromophore can therefore be suitably excited both by visible and UV radiation even though the former is the most relevant channel for photovoltaic applications.

Next we have computed the optical absorption spectrum of the precursor molecule used for the synthesis plan of Figure 1b, whose structure can be obtained by replacing one hydrogen atom of the methyl group in the diaminoterephthalate by an aromatic hydrocarbon. In order to compare the computed absorption cross section with the experimental data we have detached the precursor structure from the diad and optimized its geometry from scratch in the gas phase. The structure is shown in **Figure 3** together with the spectraum.

The computed absorption confirms that the two-band absorption scenarios described for the chromophore is unaffected by the presence of this bridge in the ground state. In the UV region of the spectrum the LR-TDDFT calculations at the B3LYP level predict absorption maxima superimposed with the measured resonances. For the optical absorption onset we obtain a peak centered at 469 nm (2.64 eV) which is in very good concordance with the experimental value of 486 nm (2.55 eV), as reported in Figure 1b and seen in Figure 3 (dashed line).

As we have mentioned above, absorption spectra can also be obtained with TDDFT beyond the linear response regime by propagating in time the Kohn-Sham system from its ground state, in presence of a wide band electric dipole perturbation. In comparison to the linear response approach, this method has the advantage to avoid explicit reference to unoccupied states. In fact the only orbitals that need to be propagated are those that are already occupied in the initial state. At this stage, the nuclei are clamped at their equilibrium positions as their motion is not crucial to determine the ground state optical absorption.

We performed calculations at the Local-Density-Approximation (LDA) level with Slater exchange[26,27] and Perdew and Zunger correlation[28] using the OCTOPUS code.[25] We remark that the width of the peaks obtained with this method is basically determined by the total time of propagation (see Section 4 for details). At present, this is not a critical issue, as we are predominately interested in the positions and relative strengths of the peaks.

We now turn our attention to the optical absorption of the full diad. Along the lines of previous works[9,10] we explore the optical response of the system at the LDA level. **Figure 4** shows the theoretical absorption spectrum of the diad, superimposed with the spectra for the chromophore, and for the acceptor (the latter is taken from Ref. [30]). The physical picture which is emerging is compatible with a scenario of weakly interacting moieties. The optical excitation can be decomposed into a visible absorption green-yellow band dominated by the chromophore absorption, and several bands in the ultraviolet having either pure acceptor character, or having contributions from both moieties. This scenario corresponds to two possible independent excitation paths, a high energy (UV) pathway, exciting both the donor and the acceptor, and the low energy (visible) one, which selectively excites the chromophore. We note that the absorption in the visible is not particularly strong and also not very broadband. Although a device built on this precise system may be expected to have a low incident photon to current conversion efficiency,[31] we may assume that the harvesting properties could be further enhanced by altering the molecular components. Therefore, we are much interested in investigating if useful charge separation mechanisms would occur. In the following section we explore the charge separation dynamics starting from a D*A type initial state.

**2.3 Charge separation**

The calculations presented in the previous section ultimately provide us with all the information we need to set up a reasonable initial state for the simulation of the photoexcited dynamical properties of the system.

In order to investigate whether or not a charge transfer can occur in the diad, we carried out unperturbed propagations of the coupled electron-nuclei dynamics according to the Ehrenfest scheme [14, 15]. We started from a Kohn-Sham excited state involving a single excitation from the chromophore HOMO to the chromophore LUMO. The calculations of the previous section have shown that this initial state is indeed accessible within LDA as a sudden change in the charge density and that it is optically active in the chromophore (almost no charge is displaced to the fullerene side).

Initially, the nuclei were given random velocities from a Boltzmann distribution corresponding to a temperature of 300 K. At our level of the theory, we do not describe energy dissipations / couplings of the system to the environment, that is, the excitation energy that was put into the system stays in the system. Thus, no relaxation channels are available for the molecule to stabilize the charge on the very long run.

The propagation was carried out with a time step of about 1 attosecond (as) for both electrons and nuclei for a propagation time of up to 85 fs. External fields were absent. In order to minimize possible spurious modulations in the time-dependent behavior of the relevant quantities, we filtered the pseudo-potentials so that they no longer contained Fourier components larger than the mesh itself [32].

The charge motion was tracked by numerical integration of the particle density in a volume including the acceptor moiety. The same evolution was computed by suppressing the electron-nuclei coupling, leaving the nuclei fixed at their equilibrium positions. This simulated experiment has proven to be very useful to clearly characterize coherent charge oscillations in other cases [9, 10].

From the structural analysis in Section 1 we concluded that a manifold of quasi-degenerate configurations exists for which the chromophore is rotated with respect to the electron acceptor. The question then arises if the *syn* and *anti* conformations, which interconvert at room temperature, might contribute differently to the charge separation dynamics, or suppress it at all. For this reason we include in the simulations an *anti*-diad chemically identical to the original *syn*-diad, but in which the chomophore is rotated by 180° around the C-C bond between the chromophore and the bridge. The structures are shown in Figure 5. While in the *syn* variant both through bonds and through space mechanism could be active, in the *anti* conformer the direct hopping probability of an electron via through space tunneling from the chromophore to the acceptor would be strongly suppressed. These may then be considered as two extreme cases within the continuum of intermediate instances mixed in a real-world sample.

**Figure 6** depicts the variation in time of the particle number on the fullerene side of both diads. Evidence of charge transfer is observed in both cases. After a 10 fs transient – witnessing quick energy relaxation and partial back bounce of the charge towards the excited moiety – the transfer from the chromophore to the acceptor sets in and proceeds unambiguously. It is also apparent from this plot that the charge transfer in the two diads proceeds differently. In the *syn* conformation (chromophore facing the acceptor) it is quantitatively more prominent, reaching an average rate of 0.014 e/fs vs 0.005 e/fs in the anti-diad. Besides, oscillations with a period of about 45 fs are clearly visible in the *anti*-conformation while for the *syn-diad* these oscillations are not exhibited due to the high transfer rate. On an absolute scale, the amount of initial charge on the acceptor in the *anti* conformation is higher than in the *syn* conformation, the corresponding electrostatic driving force is therefore smaller. The two conformations behave in a qualitatively different way, ranging from a regime of coherently driven charge transfer when the polarization is initially weak, to a regime of much stronger binding, for which the transfer occurs in an overdamped, non-oscillating regime.

As shown by the dashed lines in Figure 6, both transfer pathways are completely frozen when the nuclei are kept fixed. This implies that, in absence of structural vibrational motion, the direct tunnel probability from the donor HOMO to the acceptor LUMO is negligible, and unable to displace the charge.

## 3. Conclusions

We have thoroughly examined the structural, electronic and optical properties of a diad composed by a diaminoterephthalate molecule and a pyrrolidine-$C_{60}$ molecule bridged by a *para*-phenylene-pyrrolidine linker. Our aim was to explore the possibility of using this novel compound as a candidate for a nano-sized charge separating unit, behaving in essence as a miniature molecular solar cell. The main purpose of this investigation was to provide a simple model system, whose chemistry would be flexible enough to allow us to test several possible variants in a controlled way, and to progress towards the understanding of the charge separation process. Our theoretical results support our interest in the novel compound and call for further experimental investigations.

In summary we have found that DFT can provide a description of the chromophore in excellent agreement with the observed absorbance: the chromophore absorption onset is in the band of green light wavelength, close to the peak solar spectral irradiance. The binding of the chromophore to a well-known electron acceptor is found to be chemically stable, and its ground state level alignment is thermodynamically compatible with photoinduced charge separation. The ground state structures show that the compound is made by two basically rigid units, but a free rotation degree of freedom is present in the linking group. This fact raises the question whether two different electron transfer patterns could coexist in room-temperature samples solutions. TDDFT provides a description of the excited states of the diad as made by weakly interacting species. Two different channels for optically active excited states are observed: one corresponding to the chromophore absorption edge, the other in the UV band, corresponds to Fullerene and chromophore excitations. No optical signatures of charge transfer are observed in the ground state. We have explored the charge dynamics assuming an initial excited state corresponding to a pure chromophore HOMO-LUMO transition. From this starting point we have observed that charge is flowing from the chromophore to the Fullerene. We have also shown that the charge separation dynamics strictly depends on the relative orientations of the donor and acceptor moieties in the diad. We have evidence that the ultrafast component of the photoinduced current is related to the freedom in the nuclear motion. Nevertheless, in this case, due to the competition of different transmission channels, we do not expect that a coherent signal would be easily observable in pump-probe experiments, although we do expect that a substantial fraction of the electrons could undergo charge separation on a 100-fs time scale at room temperature. As next steps we plan to further engineer the molecule in order to improve the light-harvesting efficiency of the system in the visible region and to isolate the different transfer pathways and characterize them from both a theoretical and an experimental point of view.

## 4. Methods

In DFT the observables, in particular the ground state energy, are expressed and computed as functionals of the ground-state particle density. The single-particle orbitals of the Kohn-Sham (KS) system reproduce the particle density of the interacting system exactly – if the exact exchange-correlation functional would be known. The same orbitals are also useful to extract valuable information on the molecular orbitals and corresponding excitations. The energy differences of the KS orbitals represent the *zero-order* approximations (in the strength of the electron-electron interaction) of neutral excitations. For the actual excitations, one must to move from the DFT to TDDFT. Here the observables are expressed as functional of the time-dependent particle density, its history, and the initial state. Given that one can introduce a TD KS system, the excitations of the system may be determined in two fashions. In the first approach one works with the linear responses (LR) of the systems, as the exact interacting response function can be expressed in terms of the LR of the KS system plus corrections (that may account the effects of interaction to *all* orders). In the second approach, one may directly propagate the KS system under the effect of a perturbation to determine the time-evolved density. Real time propagation is a crucial ingredient to study charge transfer in all the cases in which the energy exchange among electrons and nuclei is important. In fact, this class of phenomena can be investigated by coupling the nuclear dynamics on the Ehrenfest path to the electron propagation in the time-dependent Kohn-Sham scheme. Here, the nuclei are evolving as classical particles under the action of the average field caused by the (quantum-mechanical) electrons and, in turn, the electrons are acted upon by the Kohn-Sham potential which includes the time-dependent potential generated by the same moving nuclei. Our analysis is restricted to the adiabatic approximation for the electrons. This means that the TDDFT exchange-correlation functionals are approximated by evaluating the DFT exchange-correlation functionals on the instantaneous density (and instantaneous single-particle orbitals – if the functionals depend also on the KS orbitals explicitly). As it is well-known, this approximation is admissible as long as memory effects do not play an important role and as long as the excitations have mostly single-particle character or collective behaviors such as plasmons. All real time propagations presented were performed with the OCTOPUS code [25], by means of a real-space discretized representation of wave functions and operators. For the chromophore we used a time step of 1.5 as and propagated up to 33 fs. For the diads we employed a time step of about 1 as and propagated up to about 85 fs. In order to reduce the required number of grid points to describe the cores of the carbon, nitrogen, and oxygen atoms, we employed Troullier-Martins pseudopotentials [29] as specified by the defaults of the code. The simulation Box was chosen to be of minimal type; i.e., each atom was contained in a sphere, whose radius was here set to 5.0 Å . The grid-spacing, in each direction, was set to 0.19 and 0.15 Å for the chromophore and for the diads, respectively.  The B3LYP level computations were performed with the GAMESS software package [33]. The LR response spectra were broadened with a Lorentzian of half-width at half-maximum of 0.2 eV.

**Acknowledgements**

SP and AD contributed equally to this work. We acknowledge financial support from the EU FP7 project "CRONOS" (grant n. 280879-2), and the PRACE Infrastructure for access to supercomputer resources at CINECA, Bologna, Italy (project n. 1562 "LAIT"). CL wishes to thank the Deutsche Forschungsgemeinschaft (SPP 1391 and DFG-NSF Materials World Network), and the Korea Foundation for International Cooperation of Science and Technology (Global Research Laboratory project, K20815000003) for financial support. JR, LF, JC and CL acknowledge support from the Deutsche Forschungsgemeinschaft (GRK 1885).

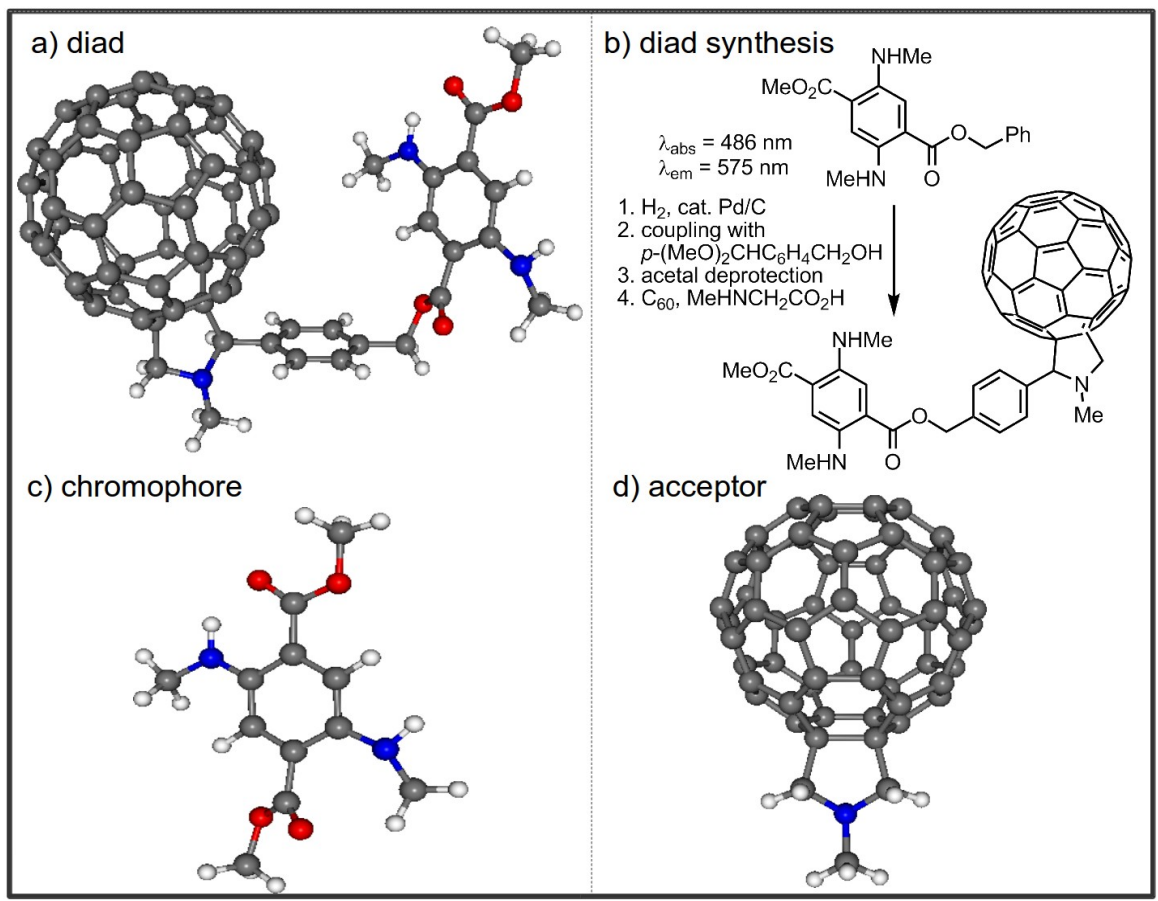

**Figure 1.** Optimized molecular structure of (a) and synthetic strategy for (b) the diad. Molecular structures of the diaminoterephthalate chromophore as dimethyl ester (donor) (c) and the pyrrolidine-$C_{60}$ (acceptor) (d). The color code for the elements is red for O, blue for N, grey for C, and white for H.

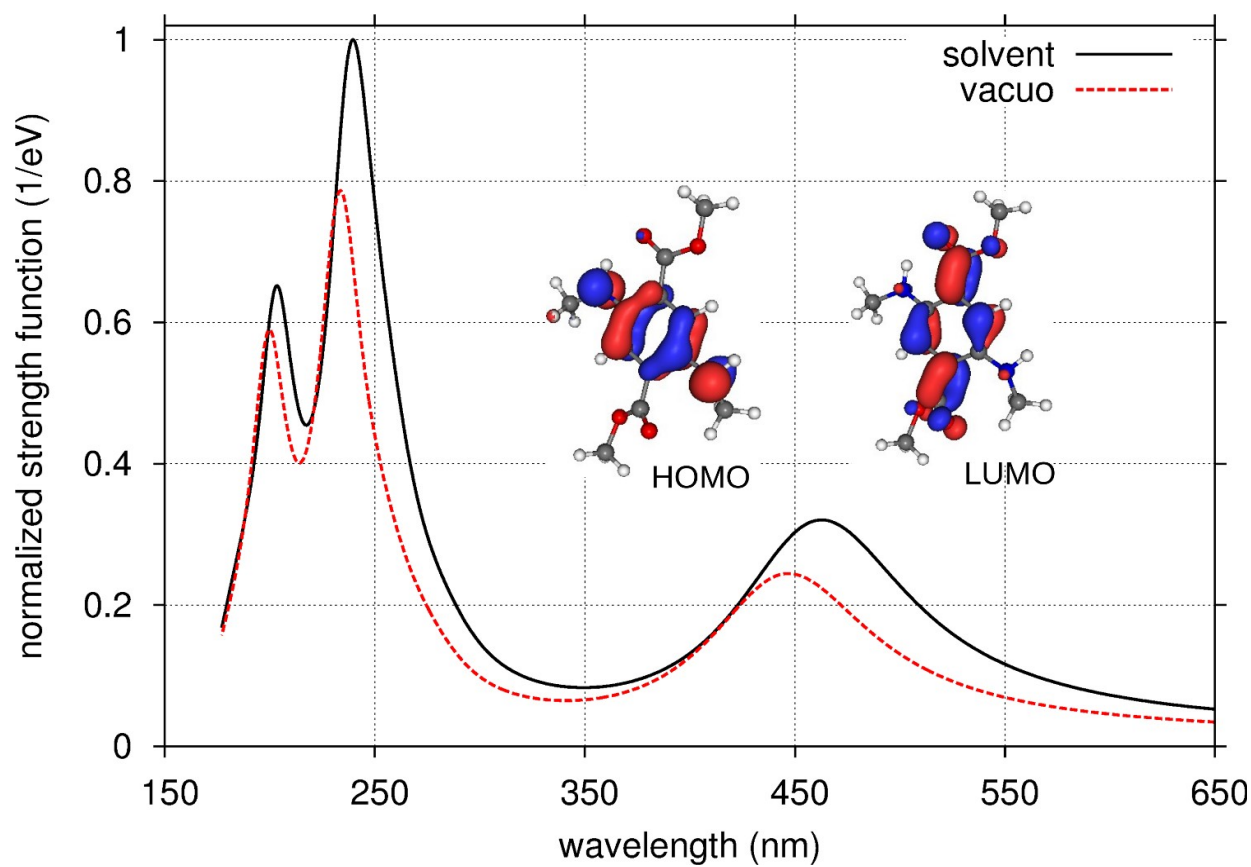

**Figure 2.** Absorption spectra of the chromophore calculated in vacuo (dash line) and in a $CH_2Cl_2$ solution (solid line) with LR-TDDFT at the level of B3LYP/6-31G(d,p). The isosurfaces of the highest occupied (HOMO) and lowest-unoccupied (LUMO) solvated molecular orbitals are superimposed. Plotted intensities are normalized respect to the absorption maximum in solvent.

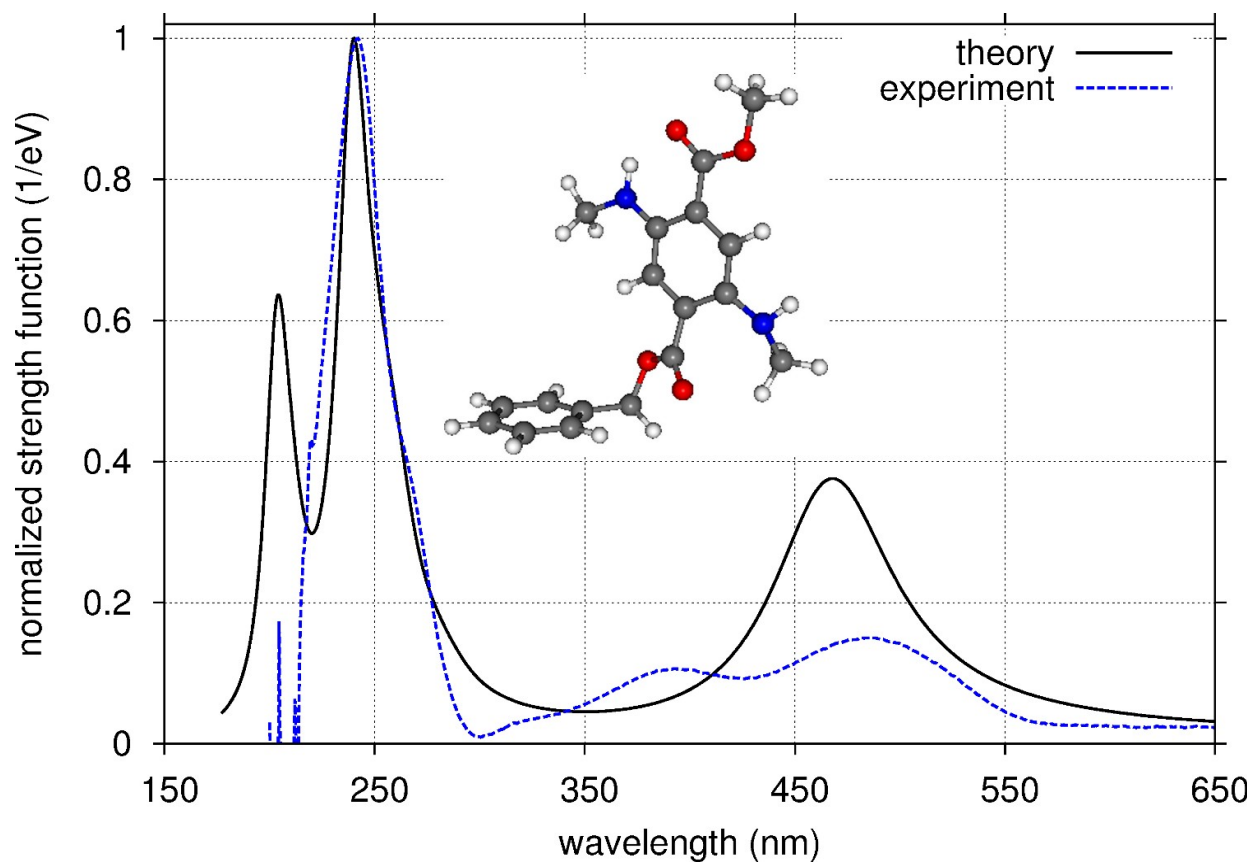

**Figure 3.** Theoretical (LR-TDDFT at the level of B3LYP/6-31G(d,p)) (solid line) and experimental (dash line) absorption spectra for the precursor molecular system in solvent ($CH_2Cl_2$). The optimized structure of the calculated molecule is superimposed.

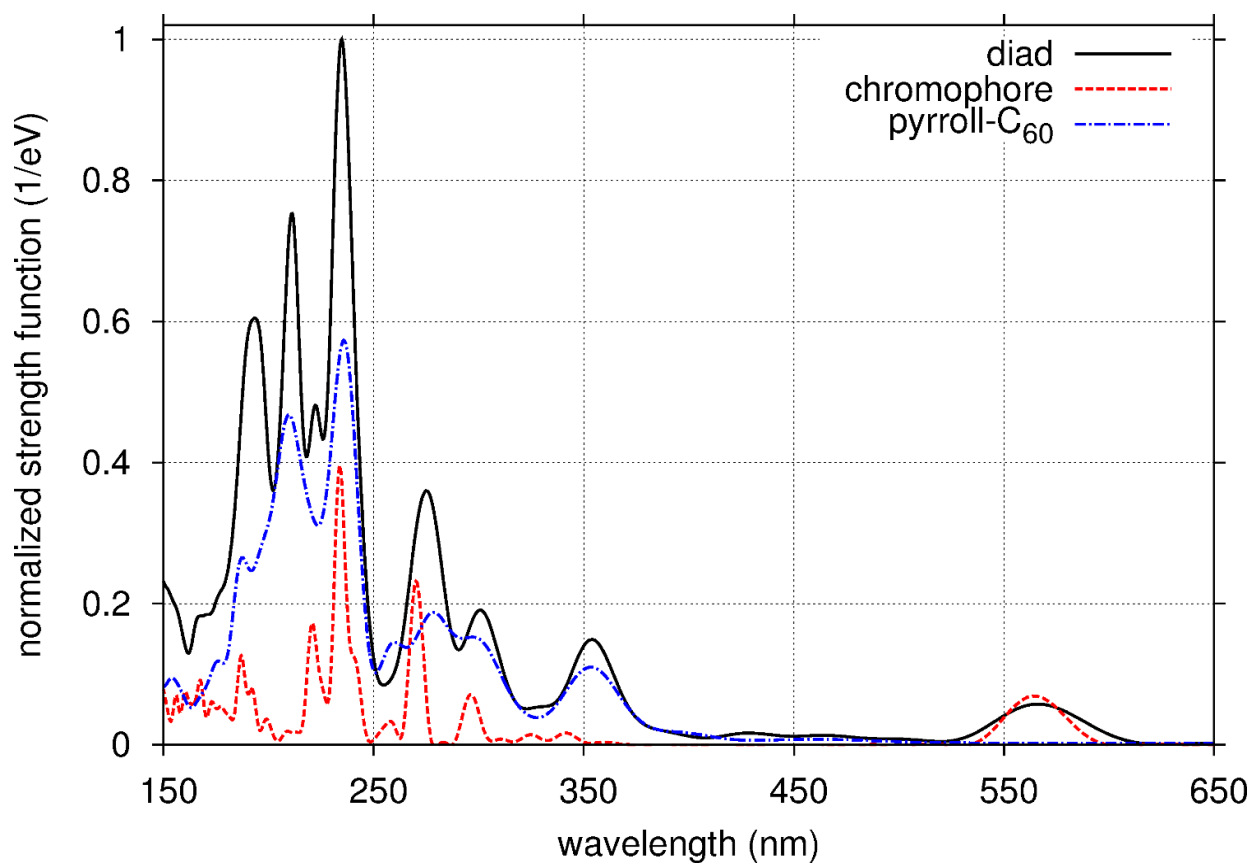

**Figure 4.** Absorption spectra of the diad (solid line), the chromophore (dash line) and the pyrrolidine-$C_{60}$ (dot-dash line) calculated by propagating in time the Kohn-Sham systems within the local density approximation. Plotted intensities are normalized with respect to the absorption maximum of the diad.

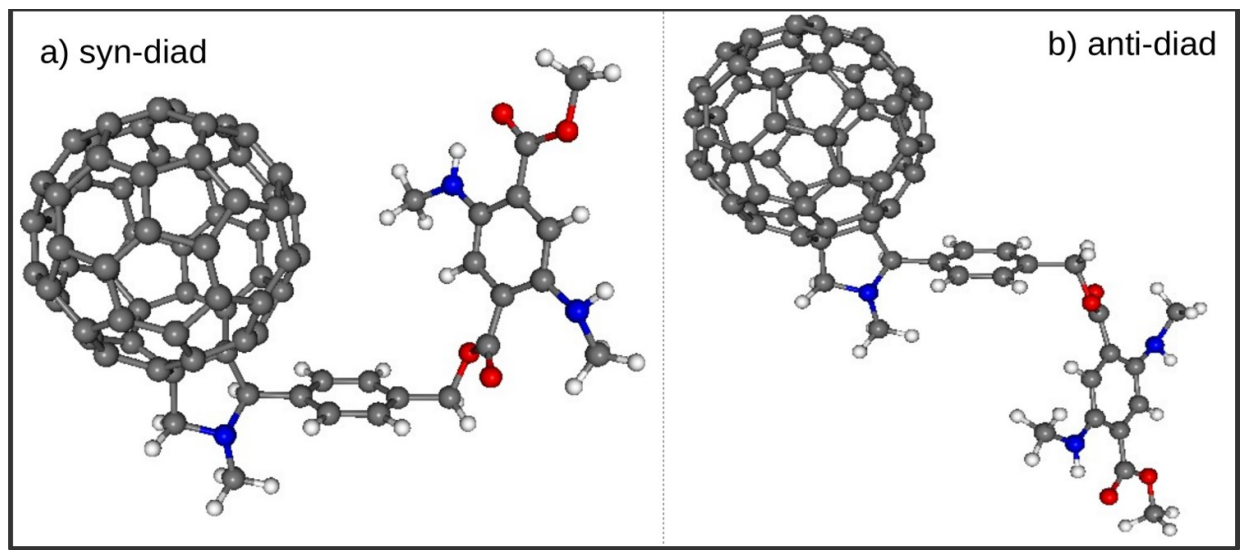

**Figure 5.** Quasi degenerate molecular structures (*syn*- and *anti*-conformations) of the diad at room temperature.

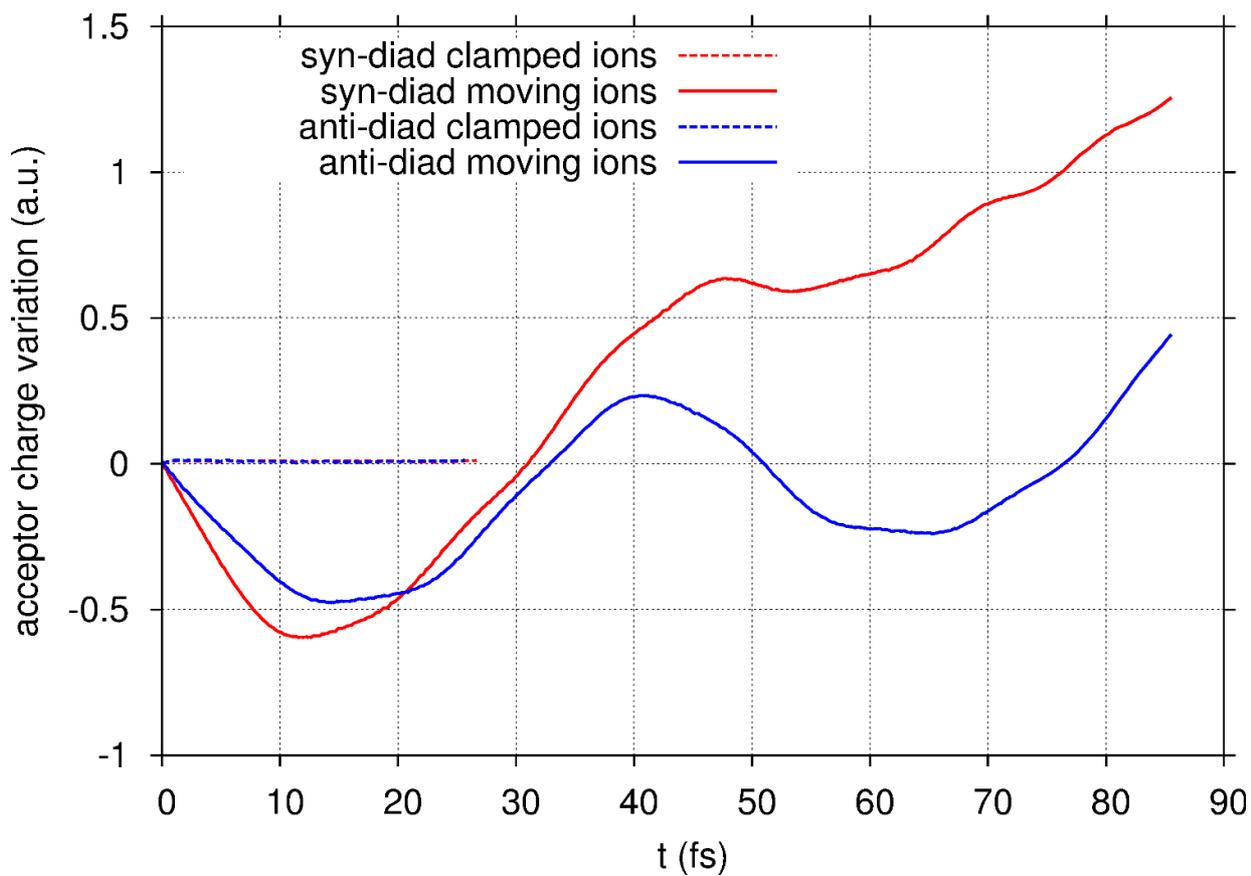

**Figure 6.** Time-dependent differential charge on the acceptor (pyrrolidine-$C_{60}$) with respect to time 0 excited state charge. Red (blue) lines refer to the *syn*-diad (*anti*-diad) depicted in Figure 5. Solid lines show the dynamics calculated with full electron-nuclei coupling (at the Ehrenfest level). Dashed lines refer to electron dynamics with clamped nuclei.